# Rational regulation strategies of interstitial localized electrons in electride: A density functional theory study

Leilei Zhang[1,2], Dan Wang[1], Hao Wang[1], Jinlong Li[1], Y. F. Wang[1], Q. Wu[1], and Hua Y. Geng[1,3,*]

[1]National Key Laboratory of Shock Wave and Detonation Physics, Institute of Fluid Physics, CAEP, P.O. Box 919-102, Mianyang, Sichuan 621900, China

[2]Henan Key Laboratory of Nanocomposites and Applications, Institute of Nanostructured Functional Materials, Huanghe Science and Technology College, Zhengzhou 450006, China

[3]HEDPS, Center for Applied Physics and Technology, and College of Engineering, Peking University, Beijing 100871, China

**Abstract**: As a class of electron-rich materials, electrides demonstrate promising applications in many fields. However, the required high pressure restricts the practical applications to some extent. This study reveals that the unique feature of electride, i.e., the localization of interstitial electrons, can be greatly enhanced and tuned by self-defective doping, applying tensile/compressive stress, or shear stress. Moreover, the requirement of orthogonality between the valence and core electron wave functions, as well as the Pauli exclusion principle, should be the driven force for the electron interstitial localization; and the exertion of external pressure modifies the available space to accommodate the electronic wave functions, thus enhances the interstitial localization. These discoveries lay down the ground for searching for promising electrides that are practicable at ambient conditions.

[*] To whom correspondence should be addressed. E-mail: s102genghy@caep.cn

**1 Introduction**

Electrides are a class of novel materials characterized by highly localized valence electrons (are often referred to as ionic interstitial quasi-atoms, ISQs) in the interstitial spaces of their crystal lattices[1-20]. Research has shown that the phenomenon of electron interstitial localization is relatively widespread in materials, particularly under compressive conditions. For example, elemental metals such as lithium (Li)[21], beryllium (Be)[22], and aluminum (Al)[23], nonmetallic elements like carbon (C)[24], as well as compounds including $SiO_3$[25], $BaMg_2$[26], and $Na_3Cl$[27], exhibit electride properties under high pressure. As a class of electron-rich materials, electrides demonstrate extraordinary lattice dynamics[22] and promising applications of this class of materials in catalysis[28], electron emitters[29], novel electrodes[30], and superconductors[31], etc. It has now evolved into an emerging and rapidly growing research hotspot. However, the required high pressure restricts the practical applications of this class of materials to some extent. In this sense, do an in-depth research on the interstitial electron localization behavior in electrides, and find an effective approach to design ambient-pressure electrides is very important for this class of emerging materials.

Lithium is the first and the simplest elemental metal in the periodic table at ambient conditions, which crystallizes in body-centered cubic (BCC) structure. It transforms into close-packed face-centered cubic (FCC) structure when cooled to zero temperature through metastable intermediate phases (including the phases of nine-layer rhombohedral (9R), hexagonal close-packed (HCP), and their variants[32, 33]). Silvi *et al*[34] once reported that the phases of Li between 0-60 GPa have interstitial-localized electrons when characterized by electron localization function (ELF), which means ambient Li might be electride. Therefore, lithium stands as an ideal candidate for investigating ambient-pressure regulation strategies of interstitial electron localization behavior by using density functional theory (DFT) method.

In this work, a systematical investigation on the electronic structure of energetically stable BCC- and FCC-Li was carried out. The possibility to enhance and tune the degree of electron localization in ambient-pressure Li is demonstrated by self-interstitial point defect doping, applying tensile/compressive stress or shear stress. This theoretical consideration gives the result that obvious enhancement of the electron localization and an unexpectedly high charge state of ISQ beyond 2*e* is achieved, showing this strategy is promising for the design of novel ambient-pressure electrides.

## 2 Computational details

The DFT calculations[35, 36] were performed using the projector-augmented wave (PAW) pseudo-potential method and plane-wave basis set as implemented in Vienna Ab-initio Simulation Package (VASP)[37-39], which has been widely employed in the study of electrides[40-42], with electronic exchange-correlation functional of generalized-gradient approximation (GGA) parameterized by Perdew-Burke-Ernzerhof (PBE)[43]. Structures of the dilated state were constructed by stretching the lattice isotropically. The density range of BCC- and FCC-Li explored in this work is spanned from $\rho_{Li} = 0.14\rho_{Li}^0$ to $\rho_{Li} = 1.0\rho_{Li}^0$, with $\rho_{Li}^0$ the density of lithium at ambient pressure. The same plane-wave basis set cutoff (1000 eV) and *k*-point mesh (spacing 0.1 Å$^{-1}$) were used to ensure the reliability and comparability of the results, which are also sufficient to ensure the convergence[44]. The total energy was ensured to converge to $1\times10^{-6}$ eV/atom. The $1s^22s^1$ electrons of Li were treated as valence electrons. The ambient-pressure structures were fully relaxed until the Hellmann-Feynman forces acting on all atoms less than 0.001 eV/Å, and the total stress tensor was converged to the hydrostatic state within 0.01 GPa.

After convergence tests, a 4×4×4 supercell of the conventional cubic cell of the BCC phase (*a*=13.828 Å) and the 3×3×3 supercell of the conventional cubic cell of the FCC phase (*a*=12.957 Å) were used to construct the self-defective systems, respectively. The point defect was introduced by adding one extra Li atom to the octahedral (*O*) or tetrahedral (*T*) interstitial site. A Monkhorst-Pack (MP) mesh of 5×5×5 was employed to sample the *k*-points in the first Brillouin zone (BZ) of the defective systems. Other convergence criteria are the same as the case of conventional unit cell. The convergence test of the *k*-point mesh, supercell size, and the energy cutoff can be found in our previous work[44].

In this work, Bader proposed atom in molecules (AIM) method[45] (i.e. Bader charge analysis) is employed to characterize the charge state of atoms and ISQs. Because this method can accurately captures the charge state information of the interstitially localized electrons when compared with other methods since it divides electrons to atoms/ISQs purely based on the zero flux surfaces of charge density distribution.

## 3 Results and discussion

### 3.1 Electron localization in ambient-pressure Li

At the beginning of this work, we investigate the electron localization behavior in elemental Li under ambient pressure. To characterize the electron localization behavior faithfully, both the ELF and charge density distribution were calculated, as illustrated in Fig.1. It is evident that interstitial localized electrons appear unambiguously in both BCC-Li and FCC-Li. One of the key indicators is that the ELF values are significantly greater than 0.5 in the interstice of BCC- and FCC-Li. Another indicator is the accumulation of charge in the interstitial sites. Besides, considering the distinct difference between anionic ISQs and covalent bonds (see Fig. S1 in the Supplemental Material (SM)) we can find that the octahedral interstitial site ($O$ site) in FCC-Li that has a strong ELF greater than 0.7 (Fig.1) in fact does not correspond to ISQ. Rather, it should be a covalent bond center (having an ELF maximum, but a charge density minimum). The bonding interactions were quantitatively characterized and validated through calculated Crystal Orbital Hamiltonian Population (COHP) analysis, with the results presented in Fig. S9. The localized ISQs are actually located at the tetrahedral site ($T$ site) in both BCC-Li and FCC-Li (having a maximum for both ELF and charge density), which is often wrongly assigned in previous investigations.

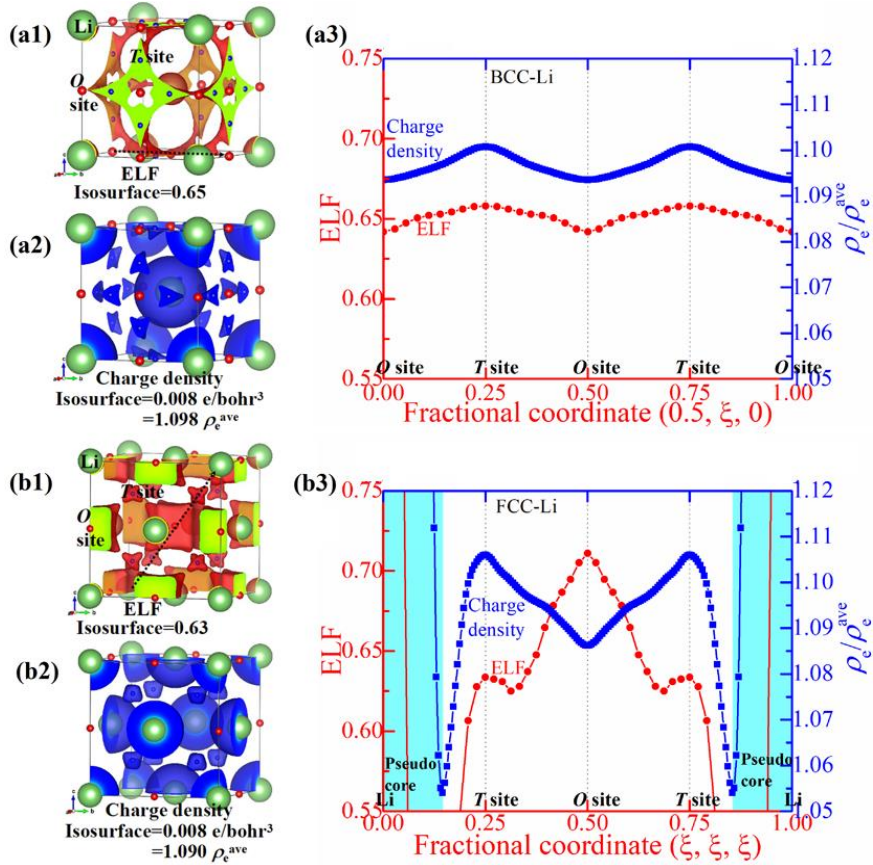

**Figure 1. (Color online)** Calculated ELF and charge density distribution in ambient-pressure Li: (a) BCC, (b) FCC, in which the isosurface values of ELF and charge density (with the average charge density $\rho_e^{ave}$ =0.0073 $e$/Bohr$^3$) are denoted below the corresponding sub-figure, and (a3) and (b3) display the variation of ELF and charge density along the given direction, respectively.

### 3.2 Electron localization from atom limit to condensed state

Then, the mechanism of the emergence of the interstitial localization of electrons was explored by tracing its evolution from the atom limit at low-density state to the condensed state at ambient-pressure. As shown in Fig. 2, as well as Fig. S10, the topology of ELF in BCC-Li changes significantly with the increasing density $\rho_{Li}/\rho_{Li}^0$. Specifically, the evolution of the electron localization can be divided into four stages: (i) All electrons are entirely attracted to localize around the nucleus and form well-defined shell structure (i.e. 'atomic limit' state). As shown in Fig. 2(a), the sharp shell distribution of the 1$s$ and 2$s$ electrons around Li is clearly visible. (ii) Electrons start to overlap between neighbouring Li atoms (i.e. the beginning of orbital overlapping), as illustrated in Fig. 2(b2). (iii) Some electrons gradually localize to the interstices, as shown in Fig. 2(c-g). (iv) Formation of highly localized anionic ISQs, as shown in Fig. 2(h). Especially, as shown in Fig. 2, the electron localization degree of the 1s orbital is almost unchanged, whereas that of the 2s orbital steadily declines as the $\rho_{Li}/\rho_{Li}^0$ increases from 0.14 to 1.00. This unequivocally shows that the interstitial electrons mainly come from the 2s orbital of Li atoms.

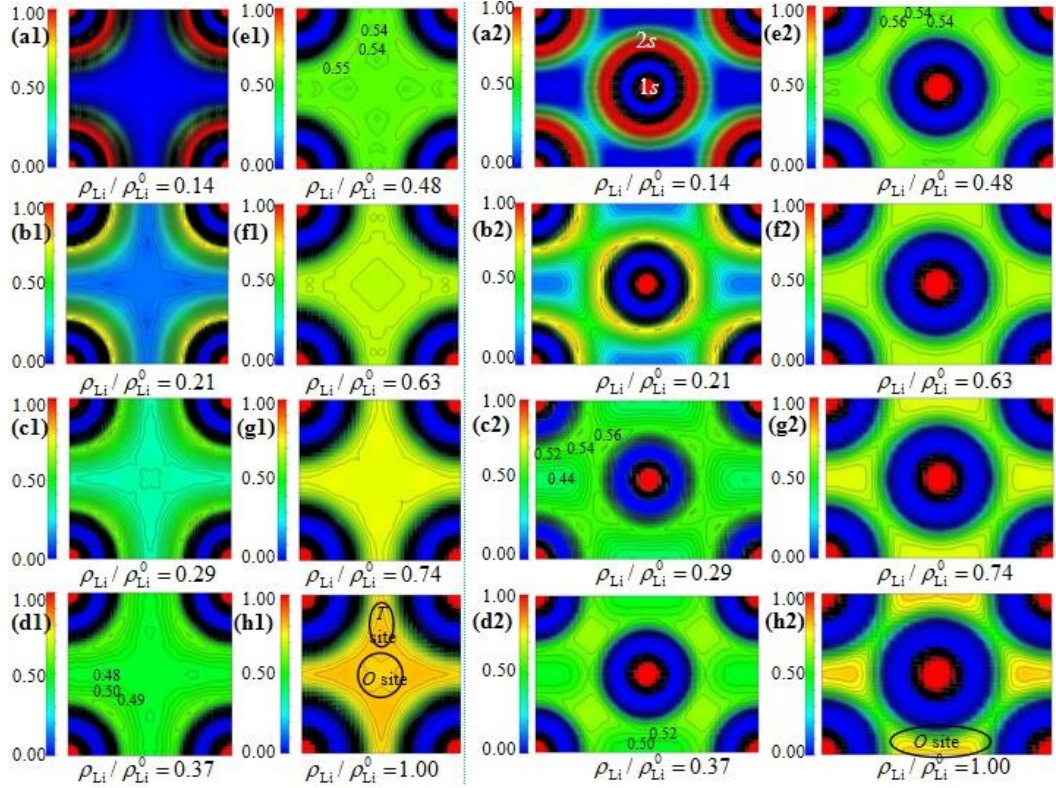

**Figure 2. (Color online)** Evolution of the calculated ELF on (a1-h1) (0 0 1) plane and (a2-h2) (1 1 0) plane in BCC-Li varying from low to high density.

The above results reveal that the requirement of orbital orthogonality between the valence and core electron wavefunctions, as well as the Pauli exclusion principle, should be the driven force for the electron interstitial localization. The exertion of external pressure modifies the available space to accommodate the electronic wavefunctions, thus alters the ordering and occupation of different orbitals (e.g., leading to s→p transition), and enhances the interstitial localization. The same conclusion holds for FCC-Li, as shown in Figs. S11-S13 in the SM. Such fundamental understanding provides critical guidance for designing ambient-condition electride materials.

## 3.3 Manipulate electron localization in ambient Li

### 3.3.1 Self-interstitial doping

Our calculations demonstrated that for both BCC and FCC-Li, the interstitial Li atom occupying the *T* site has a lower formation energy (0.698 eV and 0.745 eV) than occupying the *O* site (0.775 eV and 0.754 eV, respectively). For this reason, only the case of *T*-site defect is discussed below. As shown in Fig. 3 and Figs. S14, the introduction of interstitial Li atom enhance the electron localization degree efficiently, particularly in the vicinity of the defect site.

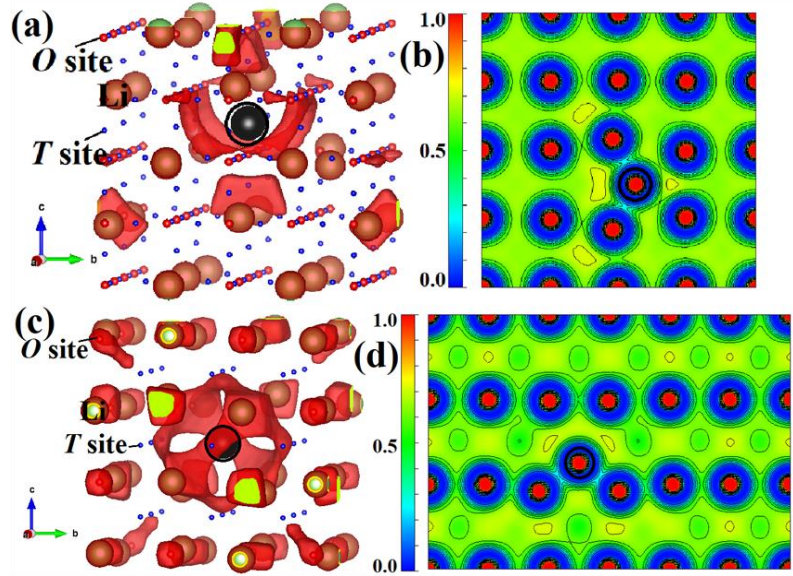

**Figure 3. (Color online)** ELF distribution with isosurface=0.68 of (a) self-defective BCC-Li and (c) self-defective FCC-Li in the vicinity of the defect site at 0 GPa; (b) and (d) show the same distribution in (0 $\bar{1}$ 0) plane of (a) and (1 $\bar{1}$ 0) plane of (c), respectively. The big green, small red, and small blue spheres represent the Li atom and octahedral interstitial (*O*) and tetrahedral interstitial (*T*) sites, respectively. The big black sphere marked with black circle in (a) and (c) represent the interstitial Li defect.

The modifications of the charge state in these two systems are also investigated, as listed in Table I. The main feature is that the charge states of *O*-site are kept being zero. Namely, the 'bonding center' nature of the *O*-site is maintained by this doping manipulation.

**Table I.** Calculated Bader charge (*Q*) of interstitial sites in defective BCC and FCC-Li. The charge difference (Δ*Q*) is with respect to the perfect BCC and FCC-Li. (Unit: *e*)

|  |  |  | NN | NNN | 3NN | 4NN | 5NN |
|---|---|---|---|---|---|---|---|
| BCC-Li | *O* sites | *Q* | 0.00 | 0.00 | 0.00 | 0.00 | 0.00 |
|  |  | Δ*Q* | 0.00 | 0.00 | 0.00 | 0.00 | 0.00 |
|  | *T* sites | *Q* | 2.49 | 0.00 | 0.00 | 0.00 | 0.00 |
|  |  | Δ*Q* | +2.36 | -0.13 | -0.13 | -0.13 | -0.13 |
| FCC-Li | *O* sites | *Q* | 0.00 | 0.00 | 0.00 | 0.00 | 0.00 |
|  |  | Δ*Q* | 0.00 | 0.00 | 0.00 | 0.00 | 0.00 |

| | | | | | | |
|---|---|---|---|---|---|---|
| *T* sites | *Q* | 0.78 | 0.00 | 1.78 | 0.20 | 0.00 |
| | Δ*Q* | +0.38 | -0.40 | +1.38 | -0.20 | -0.40 |

By contrast, the charge states of the *T*-sites are accordingly altered. For example, the amount of charge accumulation at the nearest neighboring (NN) *T*-site of the point-defect increases by 2.36*e* for BCC-Li, whereas other nearby *T*-sites decrease to null; for FCC-Li, the amount of charge at the NN and the 3-nearest neighboring (3NN) *T*-sites increases by 0.38*e* and 1.38*e*, respectively, with other *T*-sites decrease to almost null. This result demonstrates that the self-doping of Li atom can significantly enhance the local properties of electride lithium at 0 GPa. This phenomenon has also been observed in our previous studies of other systems[40, 44], further demonstrating the reliability of our results. It should be emphasized that the 4×4×4 BCC-Li supercell employed here contains 384 *O*-sites and 768 *T*-sites, and the 3×3×3 FCC-Li supercell includes 108 *O*-sites and 216-*T* sites. In other words, the point defect concentrations in BCC-Li and FCC-Li explored here are 1/1152 and 1/324, respectively. There is a large space for defective tuning to control the localization property in electride at ambient conditions. Of course, these findings require further experimental verification by techniques such as ion implantation, irradiation, and electrochemical gating.

**3.3.2 Tensile/compressive strain**

The electron localization features of BCC-Li and FCC-Li with different *c*-axis compressive/tensile strain ($\varepsilon$ =-0.15, 0.0, and +0.15, respectively) are displayed in Fig. 4. It can be seen that uniaxial compressive/tensile stress can also regulate the electride properties of ambient BCC-Li and FCC-Li. Interestingly, the ELF distribution in BCC-Li will develop into one-dimensional channel under compressive stress; whereas obvious zero-dimensional feature will be formed when the tensile stress is applied, with BCC-Li at $\varepsilon$ =0.0 as the critical point of this transition.

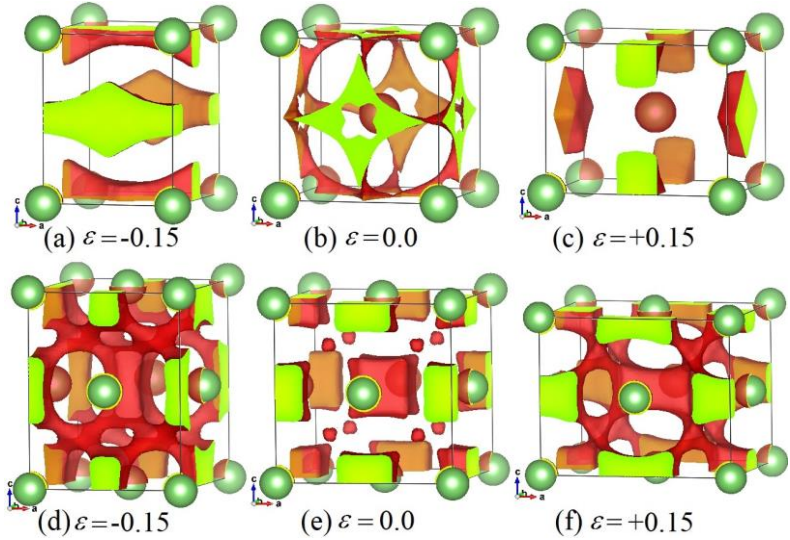

**Figure 4. (Color online)** ELF distribution of (a-c) BCC-Li with isosurface=0.65 and (d-f) FCC-Li with isosurface=0.63 with different uniaxial compressive/tensile strain along the *c*-axis, respectively.

To quantify the degree of the electron localization, a statistical analysis of the volume percentage of ELF>0.5 in BCC-Li and FCC-Li with different strain were conducted. The results are listed in Table II. It can be observed that for BCC-Li, the volume percentage of ELF in the ranges of 0.5-0.6 and 0.6-0.7 exhibit increasing and decreasing trends under uniaxial compression/tensile deformation, respectively, while the volume percentage of ELF above 0.7 remains almost unchanged. However, for FCC-Li, the volume percentage of ELF in the range of 0.5-0.6 and above 0.7 both show decreasing trends under uniaxial compression/tensile deformation, whereas the volume percentage of ELF in the range of 0.6-0.7 increases significantly. For example, at $\varepsilon$ =0.15, the volume percentage of ELF in the 0.6-0.7 range increases by 15.35% compared to that at $\varepsilon$ =0.00.

**Table II.** Calculated volume percentage of ELF>0.5 in BCC-Li and FCC-Li at different strain $\varepsilon$. (Unit: %)

|  |  | $\varepsilon$=-0.15 | $\varepsilon$=-0.10 | $\varepsilon$=-0.05 | $\varepsilon$=0.00 | $\varepsilon$=0.05 | $\varepsilon$=0.10 | $\varepsilon$=0.15 |
|---|---|---|---|---|---|---|---|---|
| BCC-Li | 0.5~0.6 | 24.01 | 25.39 | 23.61 | 23.31 | 23.59 | 24.26 | 25.46 |
|  | 0.6~0.7 | 30.25 | 29.42 | 31.45 | 31.76 | 31.29 | 30.84 | 29.86 |
|  | >0.7 | 0.71 | 0.74 | 0.73 | 0.72 | 0.74 | 0.73 | 0.72 |
| FCC-Li | 0.5~0.6 | 23.08 | 24.52 | 25.79 | 26.01 | 26.03 | 24.20 | 21.65 |
|  | 0.6~0.7 | 30.76 | 29.55 | 27.94 | 27.58 | 28.38 | 29.83 | 31.78 |
|  | >0.7 | 0.91 | 0.90 | 1.17 | 1.35 | 1.18 | 0.90 | 0.92 |

Besides, the variation of the charge state in these two systems with uniaxial deformation was also investigated. The results indicate that only the charge state of ISQs are altered accordingly, as shown in Fig. 5. In BCC-Li, the charges accumulated to the midpoint of *a*-axis increase from zero to 0.33*e* when $\varepsilon$ =0.15. In FCC-Li, the same strain engineering reduces the charge state of ISQ at *T* sites, with the electrons being transferred to *O* sites. Namely, a strain-induced electron migration between *O* sites and *T* sites in FCC-Li is predicted when $|\varepsilon|>0.10$.

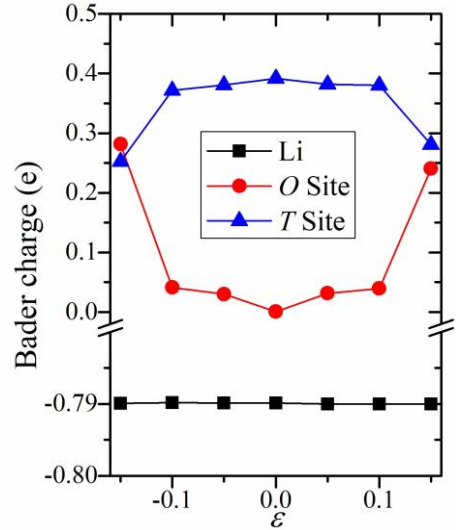

**Figure 5. (Color online)** Calculated Bader charge of atom/ISQ in FCC-Li as a function of uniaxial compressive/tensile strain $\varepsilon$. Note that the total number of *T* sites is the double of that of *O* sites.

### 3.3.3 Shear strain

The manipulation effect of the shear strain along the *c*-axis for BCC-Li and FCC-Li at ambient conditions was also investigated. As shown in Fig. 6, the ELF between *O*-sites in FCC-Li connects together and forms one-dimensional feature under shear deformation. Quantitative analysis reveals that the electron localization degree is effectively enhanced by shear stress in both BCC-Li and FCC-Li, especially the volume percentage of ELF > 0.6, as listed in Table III.

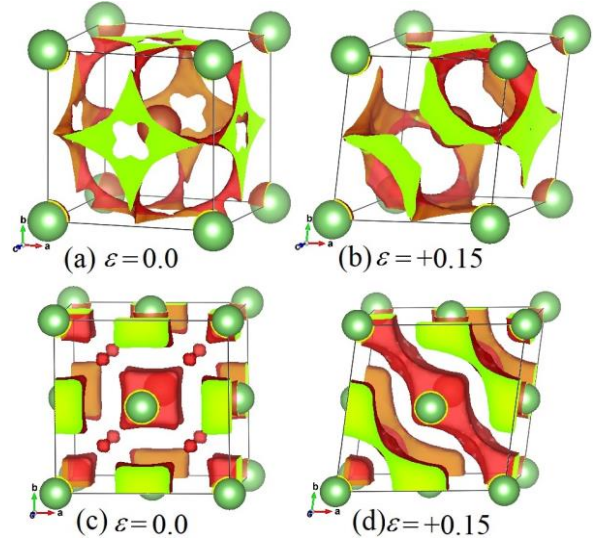

**Figure 6. (Color online)** ELF distribution of (a,b) BCC-Li with isosurface=0.65 and (c,d) FCC-Li with isosurface=0.63 at the given shear strain along the *c*-axis.

**Table III.** Calculated volume percentage of ELF>0.5 in BCC-Li and FCC-Li at different shear strain $\varepsilon$. (Unit: %)

|  |  | $\varepsilon$=0.00 | $\varepsilon$=0.05 | $\varepsilon$=0.10 | $\varepsilon$=0.15 |
|---|---|---|---|---|---|
|  | 0.5~0.6 | 23.31 | 23.41 | 22.01 | 19.72 |
| BCC-Li | 0.6~0.7 | 31.76 | 31.59 | 32.40 | 33.67 |
|  | >0.7 | 0.72 | 0.74 | 0.76 | 0.79 |
|  | 0.5~0.6 | 26.01 | 25.69 | 24.25 | 23.72 |
| FCC-Li | 0.6~0.7 | 27.58 | 27.84 | 28.58 | 28.85 |
|  | >0.7 | 1.35 | 1.47 | 1.70 | 1.88 |

Furthermore, the charge state of Li atoms in these two systems is not affected by shear strain. It also has little impact on the charge accumulation in the interstitial sites in BCC-Li, with no charge in *O*-sites and ~0.1*e* in *T*-sites. However, the charge state in interstitial sites of FCC-Li changes significantly, as shown in Fig. 7: there is an evident charge transfer from *T*-sites to the *O*-sites induced by shear strain. This indicates that the location of electron accumulation in electride can be tuned via shear deformation.

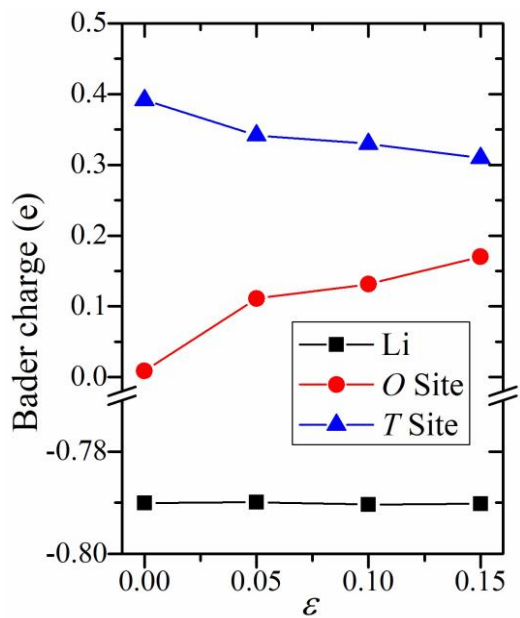

**Figure 7. (Color online)** Calculated Bader charge of atom/ISQ in FCC-Li as a function of shear strain $\varepsilon$.

Although the effectiveness of these strategies originated from studies on three-dimensional Li structures, they also provide valuable insights for low-dimensional electride materials. For instance, doping can be employed to modulate the local electron localization characteristics of low-dimensional electrides (e.g., $Ca_2N$), thereby regulating their catalytic activity. The validity of these strategies further offers new perspectives for the large-scale design of novel electride materials through machine learning and high-throughput computational approaches.

**4 Conclusion**

In summary, the electron accumulation and localization behavior in metal Li from low-density to ambient-pressure state were explored thoroughly. It was demonstrated that the electron localization process in this element (and other electride as well) has four stages: (i) electrons are completely localized to the nucleus; (ii) electronic orbitals between neighbouring atoms start overlapping; (iii) the excess electrons gradually transfer to the interstices; (iv) formation of highly localized anionic ISQs. This physical picture is general, which not only highlights the formation mechanism of electron interstitial localization, but also is intuitive to understand the complex behavior of compressed electrides. Based on this observation, we proposed a method to manipulate the ISQ and electron localization in ambient-pressure Li. A high charge state of ISQ beyond $2e$ in BCC-Li (meaning the $2s$ orbital is partially occupied), and the strain-induced charge migration between interstitial *T*-sites and *O*-sites in FCC-Li were predicted. The successful application of this approach suggests that it might be a good strategy for the design of novel ambient-pressure electrides. Especially, The finding pioneers an innovative strategy combining machine learning-based screening and high-throughput calculations for systematic development of stable electride materials.

**Supplementary Material**

See supplementary material for the detailed information about difference between anionic ISQs and covalent bonds; influence of different pseudo-potentials; linear superposition of atomic charge densities; ELF and charge density of ambient-pressure Li; calculated *vs* experimental *P-V* data of lithium; Bader charge analysis and pDOS of ambient-pressure Li; the variation of ELF and DOS in BCC-Li; Bader charge variation in low density Li; electron localization progress in FCC-Li; and strain-stress curves of ambient BCC and FCC-Li.

**Data availability**

All data needed to evaluate the conclusions in the paper are present in the paper and/or the Supplementary materials.

**Acknowledgements**

This work was supported by the National Key R&D program of China under Grant No. 2021YFB3802300, the NSAF under Grant No. U1730248, the Natural Science Foundation of Henan Province under Grant No. 242300421689, and the Specific Fund Program for Basic and Applied Basic Research of Zhengzhou City under Grant No. ZZSZX202416. Part of the computation was performed using the supercomputer at the center for Computational Materials Science (CCMS) of the Institute for Materials Research (IMR) at Tohoku University, Japan.

# Supplementary Material

# Rational regulation strategies of interstitial localized electrons in electride: A density functional theory study

Leilei Zhang[1,2], Dan Wang[1], Hao Wang[1], Jinlong Li[1], Y. F. Wang[1], Q. Wu[1], and Hua Y. Geng[1,3,*]

[1]*National Key Laboratory of Shock Wave and Detonation Physics, Institute of Fluid Physics, CAEP, P.O. Box 919-102, Mianyang, Sichuan 621900, China*

[2]*Henan Key Laboratory of Nanocomposites and Applications, Institute of Nanostructured Functional Materials, Huanghe Science and Technology College, Zhengzhou 450006, China*

[3]*HEDPS, Center for Applied Physics and Technology, and College of Engineering, Peking University, Beijing 100871, China*

## Table of contents

[*] To whom correspondence should be addressed. E-mail: s102genghy@caep.cn

## S1. Difference between anionic ISQs and covalent bonds

ELF is an important quantity to characterize the real space distribution of electron localization, which is defined as[1]

$$\text{ELF} = \left(1 + \left(D_\sigma/D_\sigma^0\right)^2\right)^{-1} \quad \quad (S1)$$

where $D_\sigma = \tau_\sigma - \frac{1}{4}\frac{(\nabla \rho_\sigma)^2}{\rho_\sigma}$ and $D_\sigma^0 = \frac{3}{5}(6\pi^2)^{2/3}\rho_\sigma^{5/3}$ ($\rho_\sigma$ is the density of electrons with $\sigma$ spin given by Kohn-Sham approximation). The $\tau_\sigma = \sum_i^N \left|\nabla \phi_i^\sigma\right|^2$ is the positive-define kinetic energy density, the Taylor expansion term $\frac{1}{4}\frac{(\nabla \rho_\sigma)^2}{\rho_\sigma}$ is the boson kinetic energy density for a system with density $\rho_\sigma$ as well as the minimum kinetic energy density that a fermion system can have (usually be called von-Weisäcker term in the chemistry), and the $D_\sigma^0 = \frac{3}{5}(6\pi^2)^{2/3}\rho_\sigma^{5/3}$ is the value of $D_\sigma$ for the homogeneous electron gas. In this definition, the value of ELF given by Eq. (1) should be within a range from 0.0 to 1.0, with ELF=0.5 corresponds to the electron-gas-like localization behaviour, and ELF=1.0 corresponds to the perfect localization.

It is necessary to point out that ELF alone is insufficient to distinguish anionic ISQs in electrides from the covalent bonding centers in covalent compounds[2], since both have a very similar ELF feature (all are corresponding to a maximum in ELF). On the other hand, we found the charge density distribution of them are quite different[2]. Take the electride hP4-Na and the covalent compound diamond for example, as demonstrated in Fig. S1, we can see that the bonding center always corresponds to a minimum in charge density, whereas anionic ISQ always has a maximal charge density value. This observation leads to an unequivocal criterion to distinguish anionic ISQ from covalent bonding center by combining ELF with charge state analysis[2]. The latter information can be obtained directly by, for example, the Bader charge decomposition method[3]. We call this combined criterion for electride the "ELF-charge criterion" hereafter.

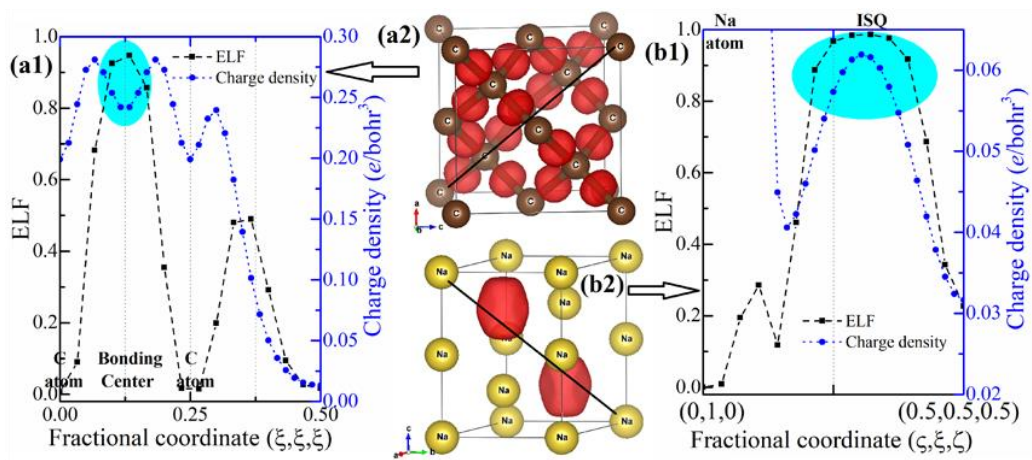

**Figure S1. (Color online)** Comparison of ELF and charge density distribution in (a1) covalent compound diamond (at 0 GPa) and (b1) electride hP4-Na (at 300 GPa); (a2) and (b2) display the crystal structures of diamond and hP4-Na and their ELF distribution (ELF=0.75), respectively. The black lines connecting the cell diagonal in (a2) and (b2) correspond to the horizontal path in (a1) and (b1), respectively. This contrast establishes the ELF-charge criterion for electrides.

Many charge analysis methods have been developed so far, such as the well known Mulliken population analysis[4], Hirshfeld population analysis[5], natural population analysis (NPA)[6], and atom in molecules (AIM) method[7]. In this work, Bader proposed AIM method[3] (i.e. Bader charge analysis) is employed to characterize the charge state of atoms and ISQs. This method can accurately captures the charge state information of the interstitially localized electrons when compared with other methods since it divides electrons to atoms/ISQs purely based on the zero flux surfaces of charge density distribution. For example, Bader charge analysis gives the charge state of the bonding center in diamond as 0.0$e$, while the given charge state of anionic ISQs in hP4-Na is 1.1$e$ (see Fig. S1). Other charge analysis methods cannot give the correct value for ISQs.

## S2. Influence of different pseudo-potentials

To estimate the influence of pseudo-potential, two different GGA-PBE PAW pseudo-potentials with different valence electrons were employed. The plane-wave basis set cutoff (1000 eV) and $k$-point mesh (spacing 0.1 Å$^{-1}$) were kept as the same, as well as the SCF convergence criterion. As listed in Table S1, the influence of pseudo-potentials on the lattice parameters and total energy is very small. Furthermore, the ELF and charge density of BCC-Li and FCC-Li calculated by using these two pseudo-potentials are shown in Fig. S2. We can see that both pseudo-potentials have quantitatively the same features, while the $1s^22s^1$ pseudo-potential gives more details since it includes all valence electrons.

**Table S1.** The lattice parameters, total energy, and total energy difference of ambient-pressure Li calculated by using two different PAW pseudo-potentials.

|  | Lattice parameters (Å) | | Total energy (eV/atom) | | ΔE with respect to FCC-Li (meV/atom) | |
| --- | --- | --- | --- | --- | --- | --- |
|  | Li: $1s^22s^1$ | Li: $2s^1$ | Li: $1s^22s^1$ | Li: $2s^1$ | Li: $1s^22s^1$ | Li: $2s^1$ |
| BCC | $a=b=c=3.433$ | $a=b=c=3.440$ | -1.905 | -1.900 | 1.520 | 1.493 |
| FCC | $a=b=c=4.322$ | $a=b=c=4.331$ | -1.907 | -1.901 | 0 | 0 |
| HCP | $a=b=3.052$ $c=5.008$ | $a=b=3.058$ $c=5.019$ | -1.906 | -1.901 | 0.272 | 0.270 |
| 9R | $a=b=3.051$ $c=22.539$ | $a=b=3.056$ $c=22.586$ | -1.907 | -1.901 | 0.052 | 0.062 |

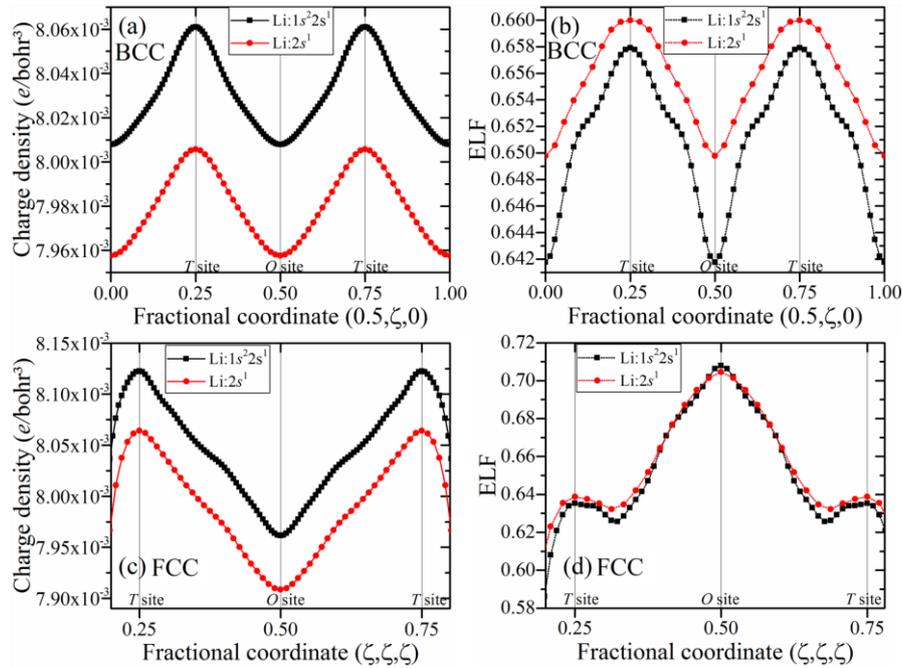

**Figure S2. (Color online)** Comparison of the ELF and charge density of ambient-pressure Li calculated with different PAW pseudo-potentials: (a) Charge density of BCC-Li, (b) ELF of BCC-Li, (c) Charge density of FCC-Li, (d) ELF of FCC-Li.

In addition, we evaluated the differences in electronic structure characteristics calculated by different functionals (PBE, PBE0, HSE03, and HSE06), as shown in Fig. S3. It can be observed that the calculated ELF and charge density profiles show negligible differences between PBE and hybrid functionals (PBE0/HSE) along the specified path. This suggests that for metallic lithium, the inclusion of exact exchange in hybrid functionals does not significantly alter the electron localization behavior or charge redistribution near interstitial regions. This can be attributed to the absence of strong electron correlation effects in lithium. However, our benchmarking reveals a substantial computational overhead (PBE: 13.5 seconds; PBE0: ~17,893 seconds (×1,320 slower); HSE03/06: ~13,000 seconds (×960 slower)).

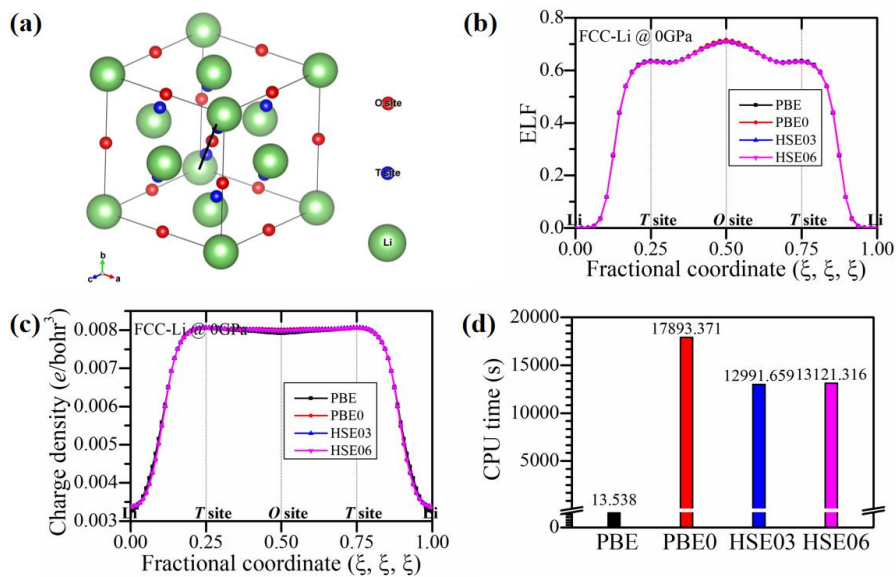

**Figure S3. (Color online)** (a) Schematic illustration of the crystal structure of FCC-Li. The large green spheres, red and small blue spheres represent Li atoms, octahedral interstitial sites ($O$ sites), and tetrahedral interstitial sites ($T$ sites), respectively. (b) Comparison of the ELF (electron localization function) distribution along the (000)-(111) direction for FCC-Li at 0 GPa, calculated using different functionals. (c) Comparison of the charge density distribution along the (000)-(111) direction for FCC-Li at 0 GPa, calculated using different functionals. (d) Comparison of CPU time consumption for static calculations of FCC-Li at 0 GPa using different functionals.

## S3. Linear superposition of atomic charge densities

Gaussian distribution for the charge density of each atom was used to simulate the effect of linear superposition of the charge densities of $s$ orbitals of two atoms when they approach to each other, as shown in Fig. S4(a). The results show that their linear superposition can not produce another maximum point of charge density between them, like what occurs in electrides. DFT calculations with two Li atoms show that the charge densities generated by the superposition of electronic orbital wave functions of each Li atoms have almost the same distribution characteristics, as shown in Fig. S4(b). This proves that the interstitial accumulation of electrons il Li can not originate from the overlapping of atomic electron clouds. Rearrangement of the atomic electron density must have been involved.

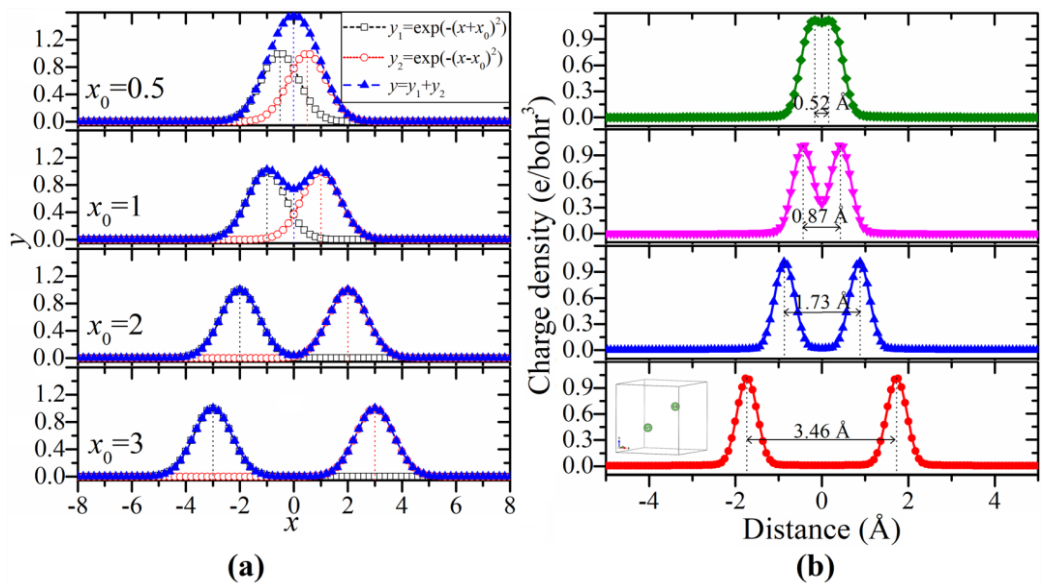

**Figure S4. (Color online)** (a) The linearly superposed charge densities of two atoms simulated by Gaussian functions; (b) The charge densities of two Li atoms calculated by non-self-consistent DFT calculation.

## S4. ELF and charge density of ambient-pressure Li

Figure S5 displays the distribution of ELF and charge density in the ambient-pressure BCC-Li and its competing phases. We can see that all energetically favoured phases (including BCC, 9R, HCP, and FCC) are in fact electride by the "ELF-charge" criterion.

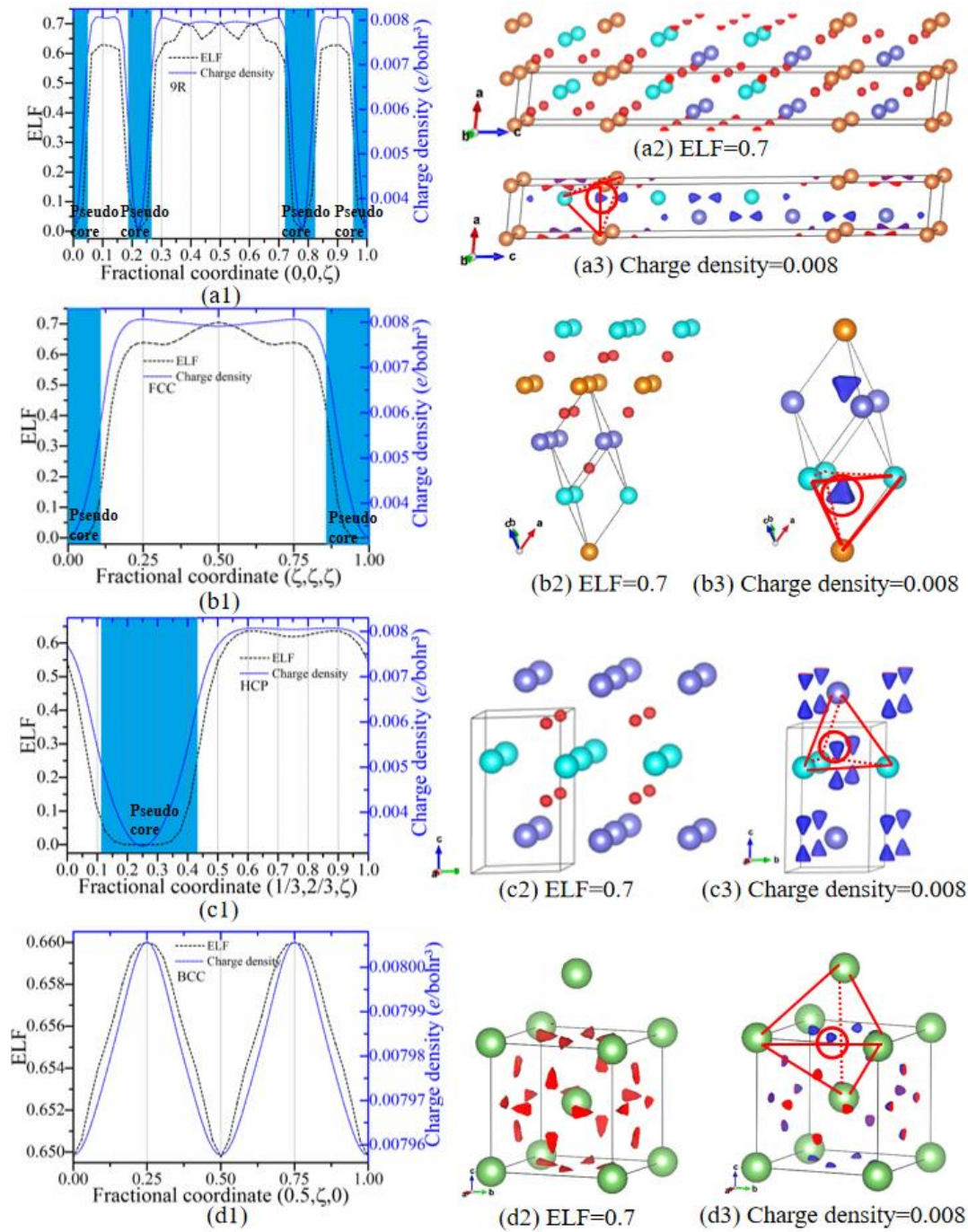

**Figure S5. (Color online)** Comparison of the ELF and charge density of ambient-pressure Li in different competing phases: (a) 9R, (b) FCC, (c) HCP, (d) BCC. The isosurface value of ELF and charge density are denoted below the sub-figures (a2)-(d2) and (a3)-(d3), respectively. The color of Li atoms in (a2)-(c2) and (a3)-(c3) are used to distinguish the different hexagonal layers. The tetrahedrons containing the interstitial localized electrons (being marked by red circles) are presented in (a3)-(d3) with red lines.

## S5. Calculated *vs* experimental *P-V* data of Lithium

Given the limited direct experimental studies on electron localization and defect behaviors in Lithium, here we have carefully compared our calculated pressure-volume (*P-V*) relationships with available experimental data. As shown in Fig. S6, our computational results demonstrate excellent agreement with experimental *P-V* measurements. This strong correlation serves as an important validation of our computational methodology and provides indirect support for our predictions regarding electron localization behavior.

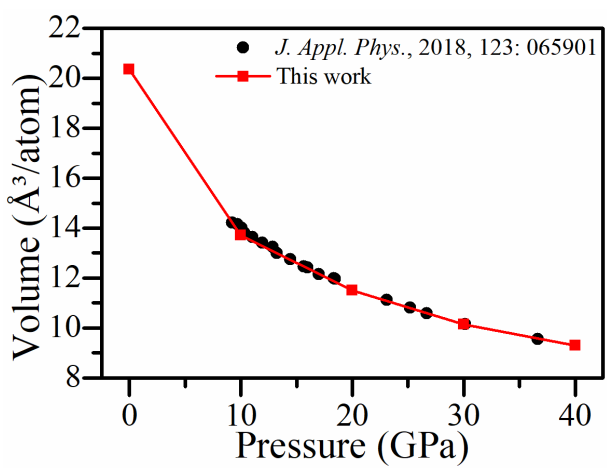

**Figure S6. (Color online)** Comparative plot of calculated versus experimental *P-V* (pressure-volume) data for lithium.

## S6. Bader charge analysis and pDOS of ambient-pressure Li

In order to characterize the electron accumulation into the interstitial *T* sites, and to explore the possible covalent bonding formation in BCC and FCC-Li, detailed Bader charge analysis and projected electronic density of state (DOS) were carried out and presented as follows.

**(1) BCC-Li**

According to the Bader charge analysis for BCC-Li, the charge state of matrix-Li atom, *O* site, and *T* site are -0.96*e*, 0.00*e*, and 0.16*e*, respectively. On the other hand, from the projected electronic DOS of BCC-Li, as shown in Fig.S7, we obtained that the number of electrons belong to the *s* and *p* orbitals of the matrix-Li atom are 2.20*e* and 0.04*e*, while the electrons projected to the *s* and *p* orbitals of the *T* sites are 0.15*e* and 0.02*e*, respectively, which agrees well with the results of the Bader charge analysis. It is thus evident that the electron transferring in ambient-pressure BCC-Li can be characterized as Li(2*s*)→ISQ$^T$. Furthermore, the weak *sp*$^2$ hybridization between Li(2*p*) and ISQ$^T$ is observed, as illustrated in Fig. S7(b).

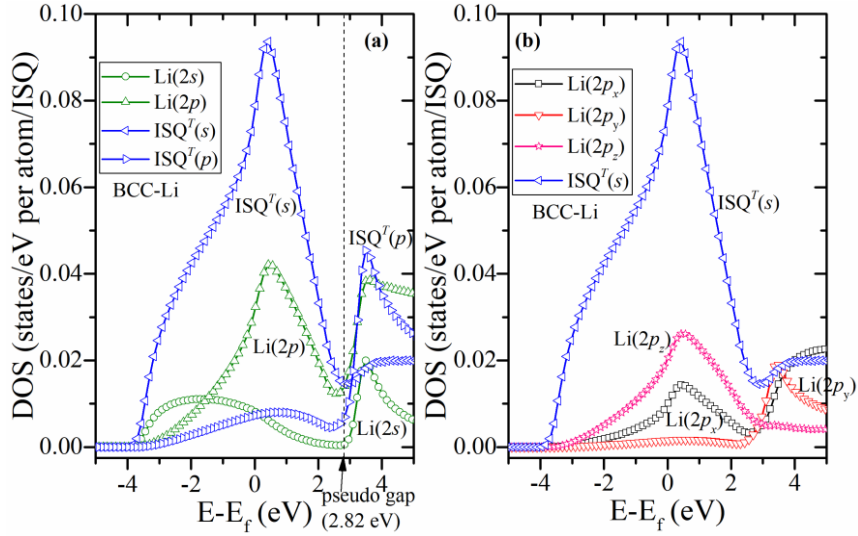

**Figure S7. (Color online)** Calculated (a) site-projected and (b) orbital-projected electronic DOS of BCC-Li at 0 GPa.

**(2) FCC-Li**

Bader charge analysis gives the charge state of the matrix-Li atom, $O$ site, and $T$ site as -0.90$e$, 0.00$e$, and 0.45$e$, respectively. Additionally, the number of electrons projected to the $s$ and $p$ orbitals of the matrix-Li atom are 2.22$e$ and 0.26$e$, and the electrons belong to the $s$ and $p$ orbitals of the $T$ sites are 0.22$e$ and 0.03$e$, respectively. Strong $sp^3$ hybridization between Li(2$p$) and ISQ$^T$ is observed (see Fig. S8(b)), which leads to a slight discrepancy between the site-projected charge state and the Bader charge analysis. For ambient-pressure FCC-Li, the electron transferring therefore can be described as Li(2$s$)→Li(2$p$)+ISQ$^T$. It is necessary to point out that we observed a pseudo-gap at about 2.8 eV above the Fermi level for both BCC-Li and FCC-Li at 0 GPa, which is formed by the splitting of the bonding and anti-bonding states due to the ISQ$^T$-Li(2$p$) hybridization.

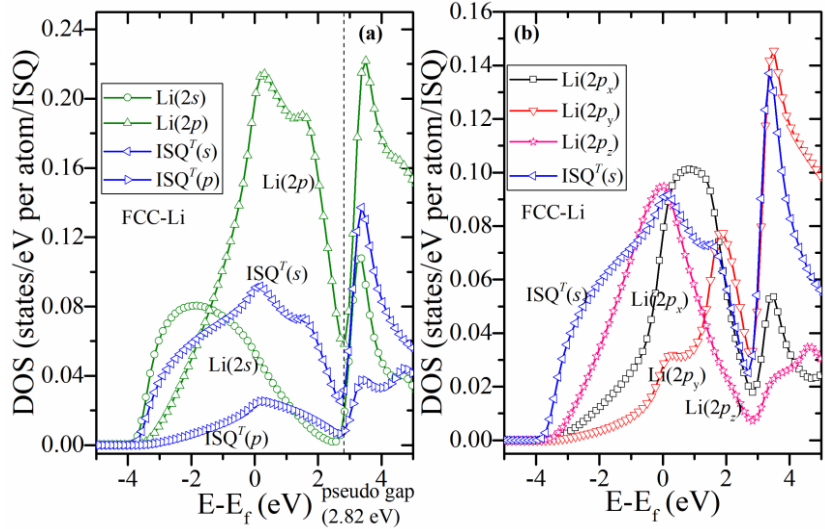

**Figure S8. (Color online)** Calculated (a) site-projected and (b) orbital-projected electronic DOS of FCC-Li at 0 GPa.

In order to measure this hybridization strength quantitatively, the Crystal Orbital Hamiltonian Population (COHP)[8] for FCC-Li was calculated using the Stuttgart version of the Tight-Binding Linear Muffin-Tin Orbitals, Atomic Sphere Approximation (TB-LMTO-ASA) program[9-11], and displayed in Fig. S9. The COHP between ISQ and ISQ/atom was calculated straightforwardly by adding an empty sphere to the ISQ position in this code. As can be seen, there are both positive and negative values for the COHP below the Fermi level between $ISQ^T$-$ISQ^T$ pair, and their ICOHP value integrated to the Fermi level is -0.014 eV/cell. It indicates that there are both bonding and anti-bonding interactions between $ISQ^T$-$ISQ^T$, but the main effect is bonding interaction. Taking the charge density distribution as depicted in Fig. S9(c) into account, we conclude that the $O$ site is the bonding center of the covalent interaction between $ISQ^T$-$ISQ^T$ pair. This gives rise to high ELF at this site. As a comparison, we also examine the chemical interactions between the nearest neighboring Li atom pair (Li-Li(NN)) and the Li-$ISQ^T$ pair. As displayed in Fig. S9, both the ICOHP of Li-Li(NN) and Li-$ISQ^T$ pairs are positive, implying that they are unfavorable in energetics if projected the electronic wavefunction back to TB-LMTO orbitals. The striking constrast with the fully relaxed and self-consistent results of orbital hybridization between $ISQ^T$ and Li atom as displayed in Figs. S8 and S9 suggests that the formation mechanism of high-pressure electride proposed by Ashcroft, namely, the claim that electrons are 'squashed into' interstitial voids[12], already breaks down even at ambient pressure. Furthermore, the accumulated electrons at the $T$ site are not corresponding to bonding center, but rather are independent anionic ISQs, which further form covalent bonds with each other with the $O$ site being the bonding center, as depicted in Fig. S9(c).

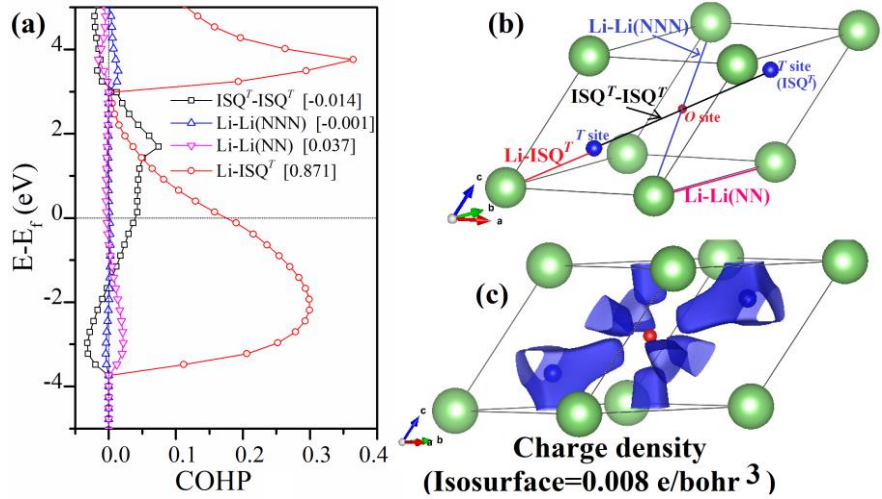

**Figure S9. (Color online)** (a) Calculated COHP of ambient-pressure FCC-Li, the value in square bracket represents the integration of COHP (i.e., ICOHP) to the Fermi level; (b) The positional relationship between the corresponding atom/ISQ pairs shown in (a); (c) Calculated charge density distribution in ambient-pressure FCC-Li.

## S7. The variation of ELF and DOS in BCC-Li

Figure S10(a-b) illustrate the variation of ELF along the lines connecting (0.5, 0.0, 0.0) to (0.5, 0.5, 0.0) and connecting (0.5, 0.5, 0.0) to (0.5, 0.5, 0.5), respectively. We can see that the electron localization degree in the interstitial space of BCC-Li increases with the increasing density. Especially, from Fig. S10(a) we can see that the $O$ site and $T$ site have almost the same electron localization degree when $\rho_{Li}/\rho_{Li}^0 < 0.55$; but $T$ site gradually builds up a local maximum with the further increase of $\rho_{Li}/\rho_{Li}^0$. From Fig. S10(b) we can see that the ELF value around the pesudo core is always around 1.00. It rapidly decreases to zero if the distance to the core is increased. This gap is almost unchanged by varying the density (see the blue area in Fig. S10(b)). As mentioned above, this area corresponds to the 1$s$ electron of Li atom. On the other hand, the ELF value near the site with a fractional coordinate of (0.5, 0.5, 0.25) steadily declines with the increasing density, and the peak completely disappears when $\rho_{Li}/\rho_{Li}^0 = 0.37$; whereas the ELF at $O$ site continuously increases from the bottom of a valley at low density and forms a peak when $\rho_{Li}/\rho_{Li}^0 > 0.55$. This variation reflects the drastic dissipation of the 2$s$ electrons. Figure S10(c)

displays the evolution of the electronic DOS around the Fermi level with $\rho_{Li}/\rho_{Li}^0$. As can be seen, when $\rho_{Li}/\rho_{Li}^0 = 1.00$, the DOS of BCC-Li agree well with the free-electron gas (FEG) model (at least in the range from the bottom of the valence band to -0.5 eV below the Fermi level). However, as the density decreases, the deviation becomes more and more pronounced.

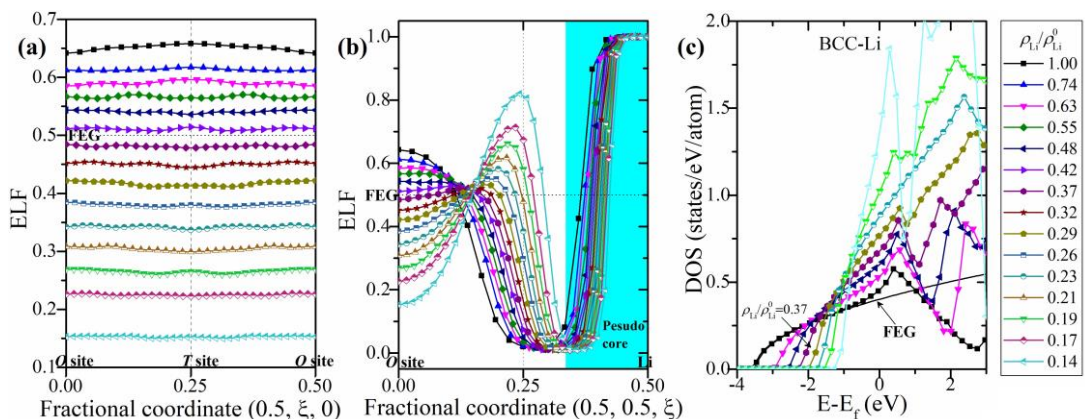

**Figure S10. (Color online)** **(a) Variation of ELF along a line from (0.5, 0.0, 0.0) to (0.5, 0.5, 0.0) in BCC-Li at different density; (b) Variation of ELF along a line from face center (0.5, 0.5, 0.0) to body center (0.5, 0.5, 0.5) in BCC-Li at different density; (c) The electronic DOS of BCC-Li for different density. The ELF and DOS of free-electron gas (FEG) model are marked as 'FEG' for comparison.**

## S8. Bader charge variation in low density Li

The Bader charge variation in BCC and FCC-Li with increasing density was investigated. The results are shown in Fig. S11. From the theory of Bader charge decomposition, we know that the 'zero' charge state for an off-atomic site means that the corresponding point locates at a minimum of the charge density distribution. Namely, no charge accumulates at that point. Therefore, with the Bader charge decomposition, we can trace the evolution of charge accumulation in the real-space, which is an important quantity supplemented to ELF to characterize electride materials. As illustrated in Fig. S11, we found that Li starts to accumulate excess charge in the interstitial sites when $\rho_{Li}/\rho_{Li}^0 > 0.6$, making the atomic charge state significantly deviates from that of a nearly-free-electron gas model. In other words, Li under ambient pressure (as well as the expanded lithium down to a density of $\rho_{Li} = 0.6\rho_{Li}^0$) is a weak metallic electride.

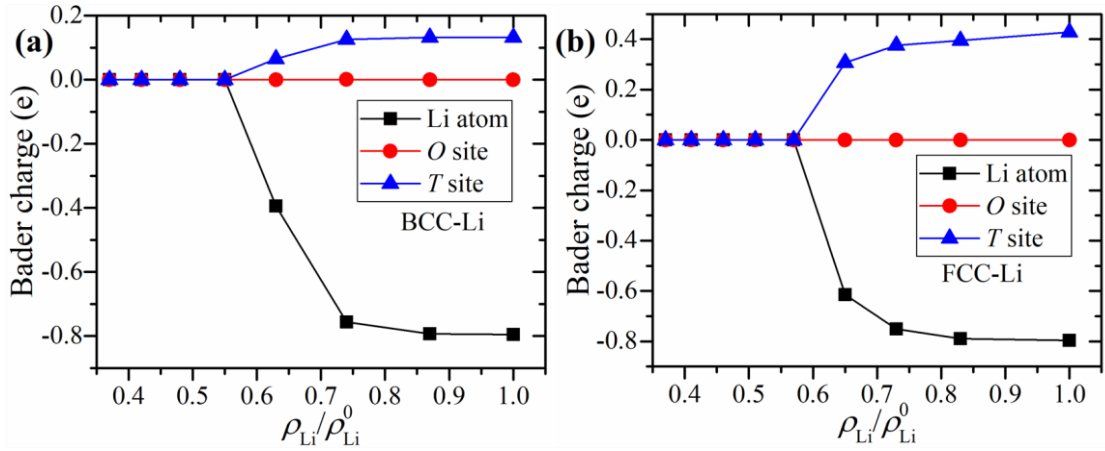

**Figure S11. (Color online)** Calculated Bader charge of atom/ISQ in (a) BCC-Li and (b) FCC-Li as a function of density ratio $\rho_{Li}/\rho_{Li}^0$, respectively.

## S9. Electron localization process in FCC-Li

Figure S12 shows the electron localization evolution process in FCC-Li with increasing the density from atom limit. It is evident that the FCC-Li has the the same localization process as BCC-Li. Namely, the localization process also can be divided into the following four stages from low density to ambient density: (i) all of the electrons are entirely attracted to around the nucleus (i.e. the 'atomic limit' state); (ii) electrons start to overlap between neighbouring atoms (i.e. the beginning of orbital overlapping); (iii) electrons gradually accumulate into the lattice interstices; (iv) formation of highly localized anionic ISQs.

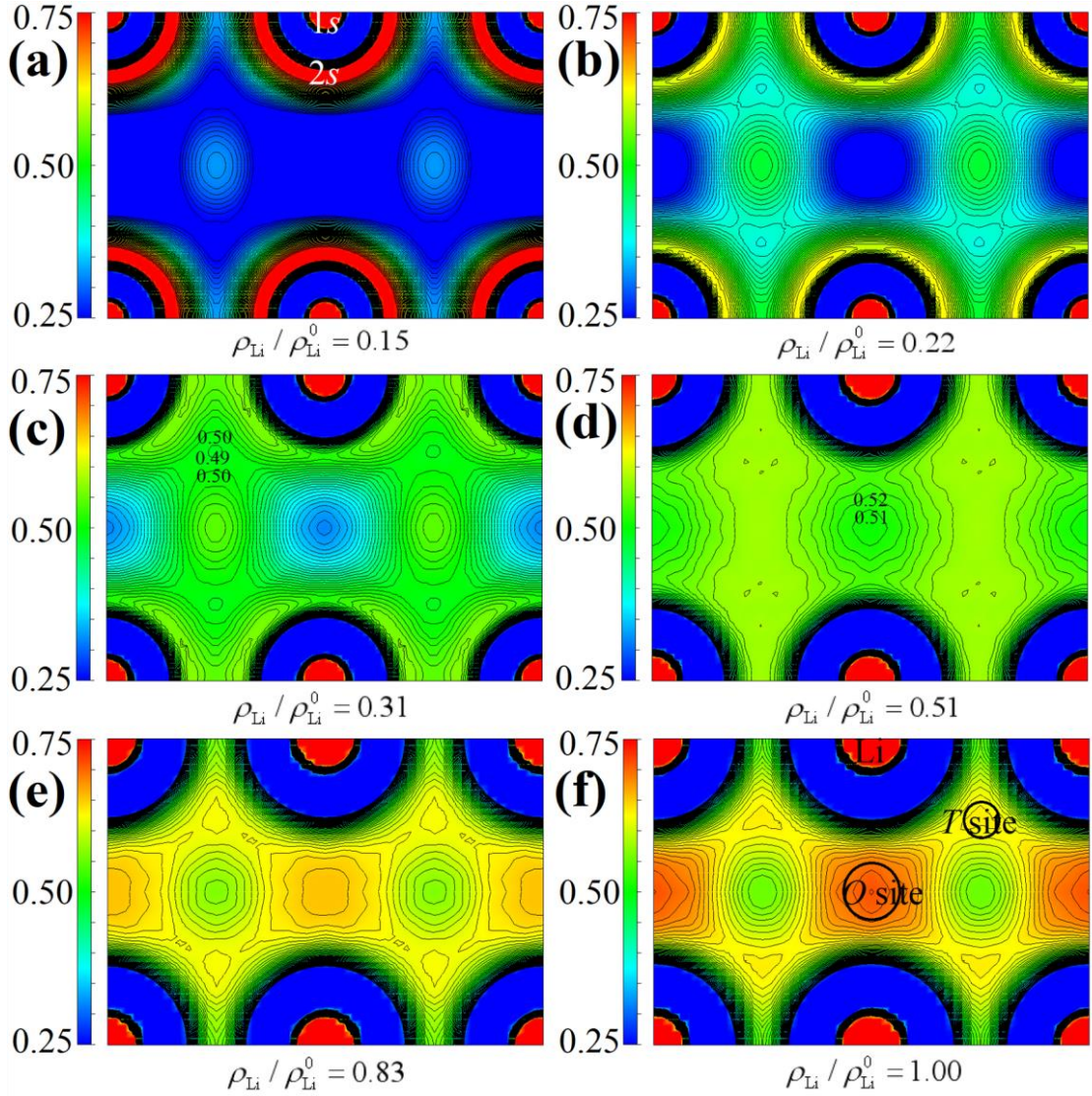

**Figure S12. (Color online) Evolution of ELF on (1 1 0) plane of FCC-Li varying from the low to high density.**

Furthermore, the electronic DOS around the Fermi level was also calculated. Figure S13 displays its variation with $\rho_{Li}/\rho_{Li}^0$. As can be seen, when $\rho_{Li}/\rho_{Li}^0=1$, the DOS of FCC-Li agrees well with the FEG model up to an energy of -0.5 eV below the Fermi level. However, as the density $\rho_{Li}/\rho_{Li}^0$ decreases, the deviation becomes more and more striking.

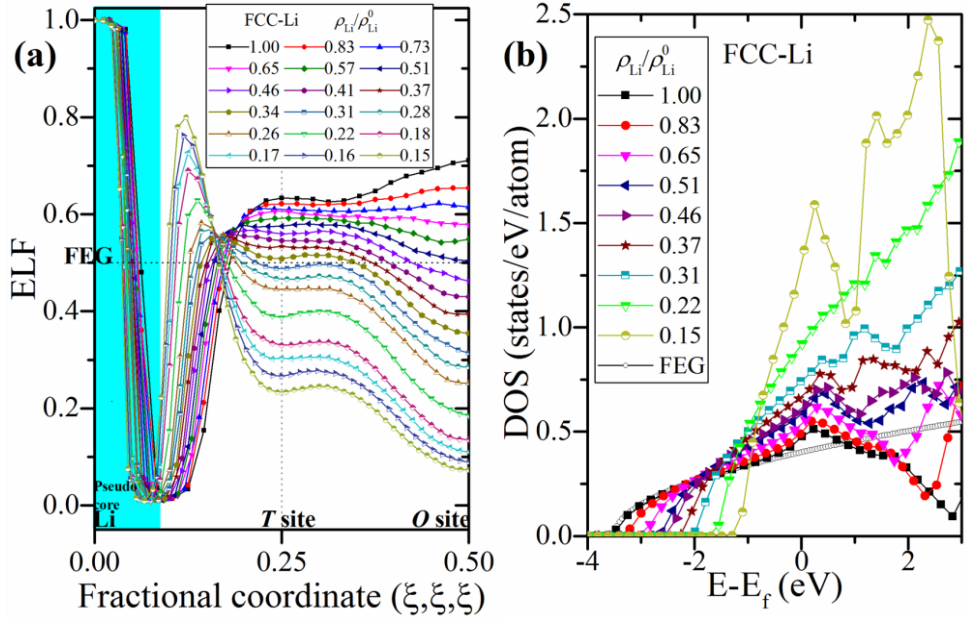

**Figure S13. (Color online)** (a) Variation of ELF along the line from (0.0, 0.0, 0.0) to the body center (0.5, 0.5, 0.5) in FCC-Li at given densities; (b) The electronic DOS of FCC-Li near the Fermi level at given densities. The results of free-electron gas model are denoted as 'FEG' for comparison. Note in (a) the 2s peak of ELF disappears at about $\rho_{Li}/\rho_{Li}^0$ =0.41.

## S10. ELF in BCC and FCC-Li

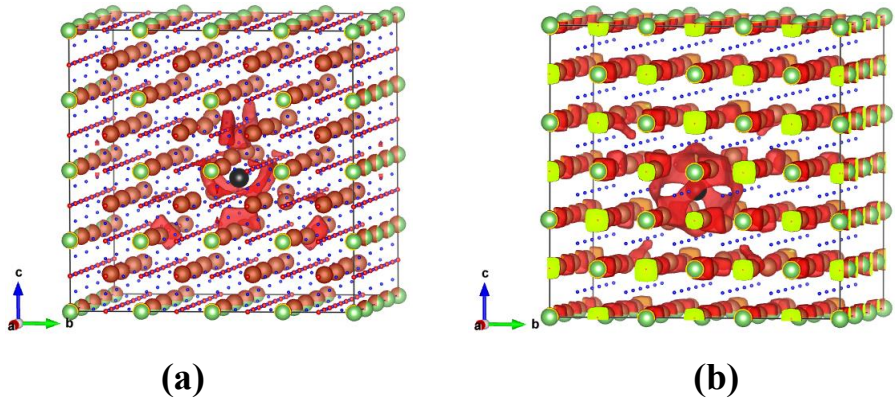

**Figure S14. (Color online)** ELF distribution with isosurface=0.68 of (a) self-defective BCC-Li and (c) self-defective FCC-Li at 0 GPa.

## S11. Stress-strain curves for ambient BCC and FCC-Li

To illustrate the magnitude of the strain as discussed in the main text, we calculated the stress-strain curves for the tensile/shear strain over a wide range, as shown in Fig. S14 and S15, respectively.

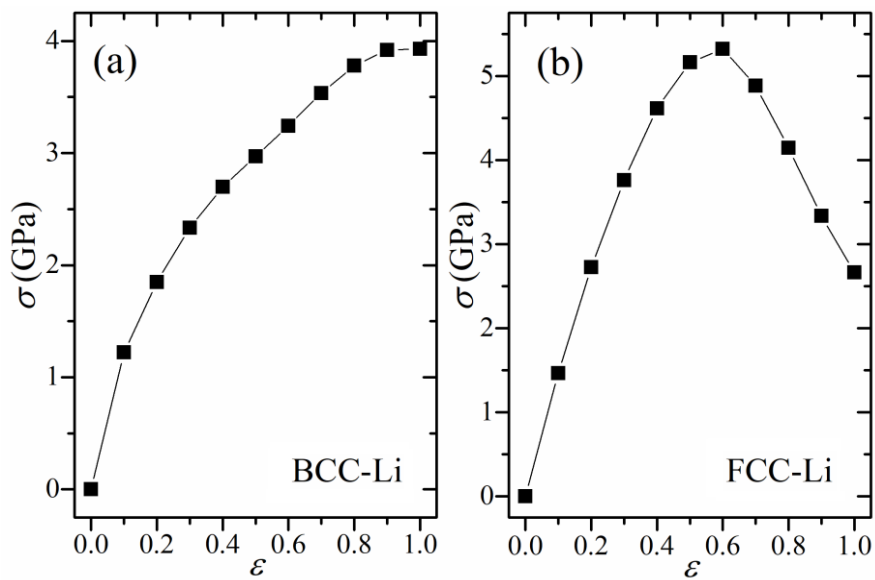

**Figure S15.** Calculated stress-strain curves of (a) BCC-Li and (b) FCC-Li at ambient pressure. The tensile strain is along the *c*-axis.

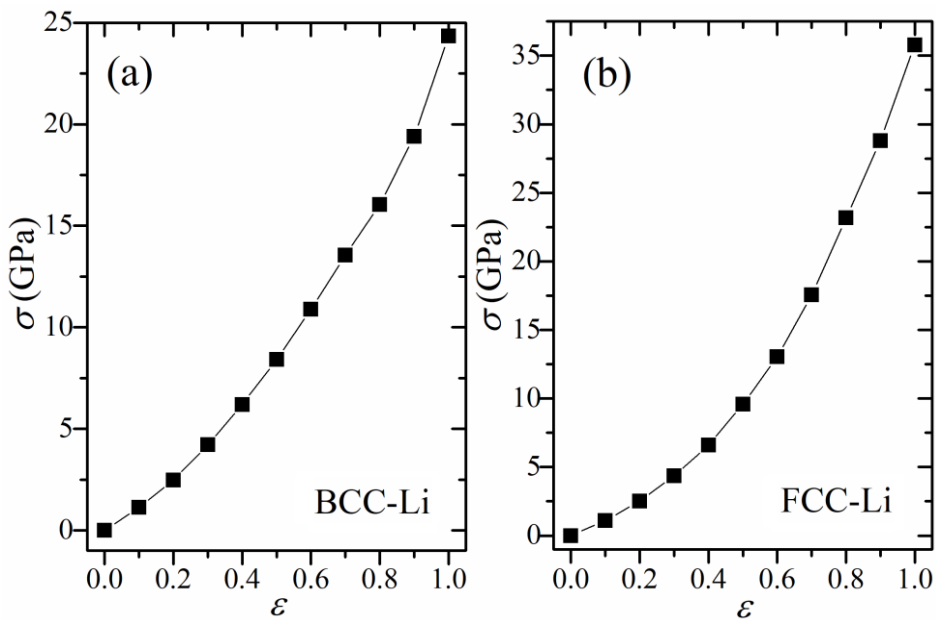

**Figure S16.** Calculated stress-strain curves of (a) BCC-Li and (b) FCC-Li at ambient pressure. The shear strain is along the *c*-axis.